\documentclass[%
  reprint,
  amsmath,amssymb,
  aps,
  prd,
]{revtex4-2}

\usepackage{graphicx}
\usepackage{dcolumn}
\usepackage{bm}
\usepackage{hyperref}

\usepackage{tikz}
\usepackage{float}
\usetikzlibrary{positioning}

\newcommand*{\labelt}[1]{\label{eqn:#1}}
\newcommand*{\flabel}[1]{}
\newcommand*{\ffig}[1]{}

\newcommand*{\eqreft}[1]{\eqref{eqn:#1}}
\newcommand*\dDp[1]{\frac{d^D#1}{(2\pi)^D}}

\newcommand*{\dba}[2]{\frac{d^{{#2}}{#1}}{(2\pi)^{#2}}}

\newcommand\teon{{(1)}}
\newcommand\tetw{{(2)}}

\newcommand\tecl{{\text{(c)}}}
\newcommand\tegf{{\text{(gf)}}}

\newcommand\teve{{\text{vertex}}}
\newcommand{\mn}{{\mu\nu}}

\newcommand{\pe}{{\phi\epsilon}}
\newcommand{\ab}{{\alpha\beta}}

\newcommand{\gd}{{\gamma\delta}}

\newcommand{\rs}{{\rho\sigma}}
\newcommand{\ez}{{\epsilon\zeta}}
\newcommand{\et}{{\eta\theta}}

\newcommand*{\abs}[1]{\lvert #1 \rvert}
\newcommand*{\absvec}[1]{\abs{\vct{#1}}}
\newcommand*{\vct}[1]{\boldsymbol{#1}}

\newcommand*{\Xmnb}[2]{\frac{\ x^\bot_{#1} x^\bot_{#2}}{\ x_\bot^2}}

\newcommand{\fcr}{\frac{\mu}{r^{\dmt}}}

\newcommand{\dmt}{{n}}

\newcommand*{\etat}[1]{\eta^{\parallel}_{#1}}
\newcommand*{\etar}[1]{\eta^{\bot}_{#1}}
\newcommand*{\hnn}[1]{{h^{(#1)}}}
\newcommand*{\hnnm}[2]{{h^{(#1)}_{#2}}}

\newcommand*{\oov}[1]{\frac{1}{#1}}
\newcommand{\ddi}{D}
\newcommand{\GNe}{{G_N}}
\newcommand{\STM}{Schwarzschild-Tangherlini}

\newcommand{\dDo}{de Donder}
\newcommand{\dDt}{\dDo-type}
\newcommand{\maP}{\mathcal{P}}

\newcommand{\maM}{\mathcal{M}}
\newcommand{\maPi}{{\mathcal{P}^{-1}}}
\newcommand{\pll}{\parallel}

\begin{document}


\title{\STM\ Metric from Scattering Amplitudes}

\author{Gustav Uhre Jakobsen}
\email{vhp485@alumni.ku.dk}
\affiliation{%
  Niels Bohr International Academy,
  The Niels Bohr Institute, University of Copenhagen
  Blegdamsvej 17, DK-2100 Copenhagen, Denmark
}%




\date{\today}

\begin{abstract}
  We present a general framework with which the \STM\ metric of a point particle in arbitrary dimensions can be derived from a scattering amplitude to all orders in the gravitational constant, $G_N$, in covariant gauge (i.e. $R_\xi$-gauge) with a generalized \dDt\  gauge function, $G_\sigma$.
  The metric is independent of the covariant gauge parameter $\xi$ and obeys the classical gauge condition $G_\sigma=0$.
  We compute the metric with the generalized gauge choice explicitly to second order in $G_N$ where gravitational self-interactions become important and these results verify the general framework to one-loop order.
  Interestingly, after generalizing to arbitrary dimension, a logarithmic dependence on the radial coordinate appears in space-time dimension $D=5$.
\end{abstract}

\maketitle


\section{Introduction}
\label{sec:intro}
The classical limit of effective quantum gravity is a successful description of general relativity.
Here, quantum field theoretic methods are used to derive results in classical general relativity \cite{Duff:1973zz,Bjerrum-Bohr:2018xdl,Goldberger:2004jt,Kosower:2018adc,Cheung:2018wkq,Bern:2019nnu,Bern:2019crd,Cristofoli:2019neg,Chung:2019yfs}.
In this approach gravitational interactions are mediated by spin-2 gravitons and general relativity is recast in the language of quantum field theory \cite{Donoghue:1995cz}.
\par
The field theoretic description of gravity is easily generalized to arbitrary space-time dimensions, $D$ \cite{Cristofoli:2020uzm,Collado:2018isu,Petrov:2020abu,Emparan:2008eg}.
Already when working with Einstein gravity in $D=4$, if the dimensional regularization scheme is used, it is to some extent necessary to work with an arbitrary dimension $D$ when pursuing the field-theoretic framework.
A classic result is the \STM\ metric which describes the gravitational field of a neutral, non-rotating point particle at rest.
\par
The quantum field description of gravity has given new insights into the gauge theory of gravity.
A well known example is the double-copy nature of gravity in terms of Yang-Mills gauge theory \cite{Bern:2010ue}.
Interest in the gauge freedom of gravity has led to the study of new perturbative gauges and field redefinitions which e.g. can be used to reduce the complexity of the Feynman rules or make apparent the double copy nature of gravity \cite{Cheung:2020gyp,Cheung:2016say}.
In general, these studies give hope that a thorough understanding, and exploitation, of the gauge freedom of gravity will result in simplifications of the complicated tensor structure of quantum gravity and possibly offer an improved starting point from which to continue investigations into quantum corrections.
\par
In this paper, we analyze the quantum field theoretic expansion of the \STM\ metric from a series of Feynman diagrams with an ever-increasing number of loops.
Such an all-order expansion was suggested in \cite{Bjerrum-Bohr:2018xdl} where it was shown how the loop integrals can be reduced in the classical limit.
Already, such expansions have been done to second \cite{BjerrumBohr:2002ks,Collado:2018isu} and third \cite{Goldberger:2004jt} order in the gravitational constant $G_N$.
\par
Our analysis uses a novel generalized gauge fixing function which combines harmonic gauge, ${g^\mn\Gamma^\sigma_\mn=0}$, and the linearized version \dDo\ gauge, $\partial_\mu h^\mu_{\sigma}=\frac{1}{2}\partial_\sigma h$.
Working all the time in arbitrary dimensions $D$ we use covariant gauge (i.e. $R_\xi$-gauge) so that our analysis depends on the arbitrary parameter $\xi$.
This approach clearly demonstrates how the classical limit depends on the quantum gauge fixing procedure.
\par
The standard coordinates of the \STM\ metric are spherical and not of the perturbative kind used in effective quantum gravity.
Perhaps the most well-known perturbative gauge is harmonic gauge.
In space-time dimension $D=4$ analytic results in harmonic gauge to all orders in $G_N$ are known \cite{Petrov:2020abu,Weinberg:1972kfs}.
However in dimensions $D\neq4$ and in \dDo\  gauge analytic results are rare.
In \cite{Goldberger:2004jt} we find the metric in \dDo\  gauge to third order in $G_N$ and in \cite{Collado:2018isu} we find it, also, in \dDo\  gauge in arbitrary dimensions $D$ to second order in $G_N$.
\par
After presenting general formulas relating the \STM\ metric to scattering amplitudes we explicitly compute the metric to second order in $G_N$.
This gives a new general result for the perturbative expansion of the \STM\ metric in the generalized gauge including both \dDo\ and harmonic gauge in arbitrary dimensions.
As a consistency check, in appendix~\ref{sec:GR} we compare the amplitude approach with a derivation using only methods from classical general relativity.
\par
In space-time dimension $D=5$ we find the curious appearance of a logarithmic dependence on the radial variable at second order in $G_N$.
This is analogous to the case in \cite{Goldberger:2004jt} in $D=4$ at third order in $G_N$.
We explain how the arbitrary scale thus introduced corresponds to a coordinate transformation which is allowed because of redundant gauge freedom.
From this explanation it is expected that the appearance of logarithmic dependence is limited to $D=5$ at second and higher orders in $G_N$ and $D=4$ at third and higher orders in $G_N$.
\par
The paper is organized as follows.
In Sec.~\ref{sec:action} we discuss the gauge-fixed action and the generalized \dDt\  gauge function in detail as well as the resulting classical equations of motion.
We consider the Feynman rules and present the graviton propagator in covariant gauge.
Then, in the first part of Sec.~\ref{sec:diagram} we consider general ideas of the all order expansion of the \STM\ metric in terms of scattering amplitudes.
In the subsections~\ref{sec:newton} and~\ref{sec:second} we compute the first and second order contribution to the metric, respectively.
In Sec.~\ref{sec:logarithm} we discuss the appearance of logarithms in the metric.
There are three appendices.
In Appendix~\ref{sec:frules} we present the relevant Feynman rules for our computations and in Appendix~\ref{loopint} we analyze the triangle integrals for the one-loop amplitude.
In Appendix~\ref{sec:GR} we go through the alternative derivation of the expansion of the \STM\ metric.
\par
\section{Covariant and Generalized Gauge Fixing}
\label{sec:action}
We work with the Einstein-Hilbert action minimally coupled to a massive scalar field together with the covariant gauge fixing term:
\begin{equation}
  S = \int d^\ddi x \sqrt{-g}
  \Big(
  \frac{2R}{\kappa^2}
  + \mathcal{L}_\phi
  \Big)
  + \int d^\ddi x
  \frac{\eta^\mn G_\mu G_\nu}{\kappa^2\xi}
  \ .
  \label{eqn:act2}
\end{equation}
Here $\kappa^2=32\pi\GNe$ and $\mathcal{L}_\phi=\frac{1}{2}(g^\mn\partial_\mu\phi\partial_\nu\phi - m^2\phi^2)$ and we use the mostly minuses metric.
Also, $\xi$ is the covariant gauge-parameter.
Additionally, from the path integral gauge fixing procedure, there would be a ghost term which we, however, will not consider since it does not contribute in the classical limit.
\par
We choose a \dDt\  family of gauge functions $G_\sigma$ which depend on the arbitrary parameter $\alpha$:
\begin{equation}
  G_\sigma =
  (1-\alpha) \ \partial_\mu (h^\mu_\sigma - \frac{1}{2} \eta^\mu_\sigma h_\nu^\nu)
  + \alpha\  g^\mn \Gamma_{\sigma\mn}
  \ .
\end{equation}
Here $h_\mn=g_\mn-\eta_\mn$  and indices on $h_\mn$ are raised and lowered with the flat space metric.
The index on $\Gamma_{\sigma\mn}$ was lowered with $g_\mn$.
\par
Note the details of this gauge function.
When $\alpha=0$ we have \dDo\  gauge,
$\partial_\mu (h^\mu_\sigma - \frac{1}{2} \eta^\mu_\sigma h)=0$
and when $\alpha=1$ we have harmonic gauge
$g^\mn \Gamma_{\sigma\mn}=0$.
Here we have used similar terminology as \cite{Cheung:2020gyp}.
Any choice of $\alpha$, however, results in a perturbatively valid gauge choice of the same generalized type as discussed in \cite{Cheung:2020gyp}.
When $G_\sigma$ is expanded in $h_\mn$ the linear term is independent of $\alpha$ while the non-linear terms are linear in $\alpha$.
Thus, the gauge parameter $\alpha$ scales all the non-linear terms of $G_\sigma$.
For the one-loop computation we need only the linear and quadratic terms which we find to be:
\begin{equation}
  G_\sigma
  =
  h^\mu_{\sigma,\mu}
  -
  \frac{1}{2}
  h_{\mu,\sigma}^\mu
  -
  \alpha\
  \Big(
  h^\mu_\nu
  h^\nu_{\sigma,\mu}
  -
  \frac{1}{2}
  h^\mu_\nu
  h^\nu_{\mu,\sigma}
  \Big)
  +\mathcal{O}(h^3)
  \ .
\end{equation}
Here, and later, we use the comma-notation for partial derivatives.
\par
The classical equations of motion $\delta S = 0$ depend on both gauge parameters.
First, we will focus on the dependence on the covariant parameter.
We get the equations of motion
\begin{subequations}
  \labelt{ext6}
  \begin{eqnarray}
    &&G^{\mn} + \frac{1}{\xi} H^{\mn} = -\frac{\ \kappa^2}{4} T^{\mn}
    \ ,
    \label{eqn:ein1}
    \\
    &&\sqrt{-g} H^\mn
    =
    \maP^{\mn\rs}
    \partial_\sigma
    G_\rho
    +
    \alpha
    G^\rho \Gamma_{\rho\ab} g^{\alpha\mu}g^{\beta\nu}
    \labelt{hte1}
    \\
    &&
    \qquad\qquad\quad
    +
    \alpha
    \Big(
    I^\mn_{\rho\kappa} I^{\sigma\kappa}_\ab
    -\frac{1}{2}
    \delta^\sigma_\rho I^\mn_\ab
    \Big)
    \partial_\sigma
    \big(
    G^\rho
    (g^\ab-\eta^\ab)
    \big)
    \nonumber
    \ ,
  \end{eqnarray}
\end{subequations}
where in Eq.~(\ref{eqn:hte1}) we use the notation $I^\mn_\ab = \frac{1}{2}(\delta^\mu_\alpha\delta^\nu_\beta + \delta^\mu_\beta\delta^\nu_\alpha)$ and $\mathcal{P}^\mn_\ab = I^\mn_\ab-\frac{1}{2}\eta^\mn\eta_\ab$ from e.g. \cite{Donoghue:1995cz,BjerrumBohr:2002kt} and indices on $G_\sigma$ and $\maP_\ab^\mn$ are raised with the flat space metric.
In Eq.~(\ref{eqn:ein1}), $G^\mn$ is the Einstein tensor, $T^\mn$ is the energy-momentum tensor of matter, and $H^\mn$ is a Lorentz covariant tensor which breaks the general covariance of the Einstein field equations.
\par
We get an additional equation to Eq.~\eqreft{ein1} by taking the covariant derivative on both sides.
Since both $D_\mu G^\mn$ and $D_\mu T^\mn$ vanish by themselves, the covariant derivative of the gauge-breaking tensor $H^\mn$ is forced to disappear as well:
\begin{equation}
  D_\mu H^\mn = 0
  \ .
  \label{eqn:con1}
\end{equation}
We interpret Eq.~\eqreft{con1} as a gauge condition on the metric, $g_\mn$.
\par
It is still not clear how to solve the equation of motion Eq.~\eqreft{ein1} together with the gauge condition Eq.~\eqreft{con1} and what roles the covariant parameter, $\xi$, and $H^\mn$ play.
However, it is easy to construct a metric which satisfies these equations.
Namely, if we choose a metric which obeys the Einstein field equations,
\begin{equation}
  G^{\mn} = -\frac{\ \kappa^2}{4} T^{\mn}
  \ ,
  \label{eqn:con2}
\end{equation}
together with the simple gauge condition $G_\sigma=0$.
This metric solves Eqs.~\eqreft{ein1} and~\eqreft{con1}.
This is so since from the definition of $H^\mn$ in Eq.~\eqreft{hte1} it is clear that $G_\sigma=0$ implies that $H^\mn$ vanishes.
Hence this metric trivially solves the gauge condition Eq.~\eqreft{con1} and due to the vanishing of $H^\mn$ Eq.~\eqreft{ein1} reduces to the Einstein field equations which the metric is assumed to satisfy.
\par
In a perturbative expansion of Eq.~\eqreft{ein1} where $h_\mn$ is expanded in powers of $G_N$ and $T^\mn$ is given as a source describing the point particle at rest, the propagator of $h_\mn$ is well-defined (and is the same as the quantum propagator presented below).
Hence the metric is uniquely defined by the equation of motion, Eq.~\eqreft{ein1}, to the extend that the source, $T^\mn$, is so and with the exception of possible scales from the renormalization of divergencies.
We expect this (unique) solution to be the metric $g_\mn$ discussed above which solves the Einstein field equations, Eq.~\eqreft{con2}, with the gauge condition $G_\sigma=0$.
\par
In this case the metric is clearly independent of the covariant parameter, $\xi$, since it is defined by equations which are both independent of $\xi$ and we see then, that the two gauge parameters $\xi$ and $\alpha$ play very different roles.
The covariant gauge parameter $\xi$ appears only during intermediate steps and the classical metric is independent of $\xi$.
During the calculation, however, it is convenient to separate quantities into parts according to their dependence on $\xi$.
The parameter $\alpha$ is introduced to describe an entire family of classical gauge choices.
The classical metric then depends on $\alpha$ since the gauge condition $G_\sigma=0$ does.
\par
To derive the Feynman rules we expand the action around flat space-time in $h_\mn$.
Since the linear term of the gauge function $G_\sigma$ is independent of $\alpha$, the quadratic term in the action $S$ will also be independent of $\alpha$.
From the quadratic term in $S$ we derive the graviton propagator in covariant \dDo\ gauge in momentum space:
\begin{equation}
  \frac{iG^\mn_\ab }{q^2+i\epsilon} =
  \frac{i}{q^2+i\epsilon}
  \Big(
      {\mathcal{P}^{-1}}^\mn_\ab
      - 2(1-\xi)
      I^{\mn}_{\rho\kappa} \frac{\ q^\rho  q_\sigma}{q^2} I^{\kappa \sigma}_{\ab}
      \Big)
      \label{eqn:pro1}
      \ .
\end{equation}
Here ${\mathcal{P}^{-1}}^\mn_\ab$ is the inverse operator to $\mathcal{P}^\mn_\ab$ which is the well known \dDo\ propagator
\begin{equation}
  {\mathcal{P}^{-1}}^\mn_\ab = I^\mn_\ab-\frac{1}{D-2}\eta^\mn\eta_\ab
  \ ,
  \label{eqn:ddo1}
\end{equation}
to which the covariant propagator reduces for $\xi=1$.
For other values of $\xi$ a new momentum-dependent term appears in the propagator.
Later, it will be convenient to separate the propagator into two terms, one independent of $\xi$ and the other linear in $\xi$.
\par
Expanding the action in $h_\mn$ generates terms with an arbitrary number of gravitons.
For the one-loop calculation only the $\phi^2 h$ and $h^3$ vertices are necessary.
These are included in Appendix~\ref{sec:frules}.
We note, however, how the vertices in general depend on the gauge parameters $\xi$ and $\alpha$.
The coupling of $h$ to $\phi$ is independent of the gauge fixing and hence the vertices $\phi^2 h^n$ as well.
The graviton self-interaction vertices can conveniently be separated into two terms, one independent of $\xi$ and one linear in $\oov{\xi}$.
As for the terms linear in $\oov{\xi}$ these can then be divided into terms linear or quadratic in $\alpha$.
\section{Diagram Expansion of the \STM\ Metric}
\label{sec:diagram}
\begin{figure}[h]
  \centering
  \includegraphics[width=5cm]{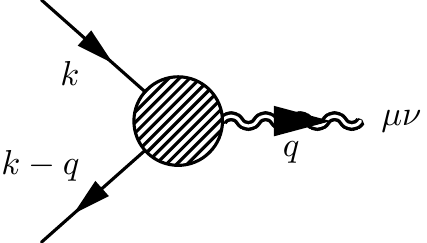}
  \caption{
    A massive scalar interacts with a graviton.
    The diagram represents the vertex function, $i\mathcal{M}_{\text{vertex}}^\mn$.
    In the classical limit, it acts as the source of the metric.}
  \label{fig:amplitude1}
\end{figure}

It is an exciting idea that the \STM\ metric can be computed from scattering amplitudes and Feynman diagrams \cite{Duff:1973zz,Bjerrum-Bohr:2018xdl}.
Since the metric is not a gauge-invariant object, the relevant diagrams cannot be gauge-invariant either and they will include an external graviton.
In this section we relate the \STM\ metric to the vertex function of a massive scalar interacting with a graviton.
This amplitude is shown in Fig.~\ref{fig:amplitude1}.
In the classical limit, diagrams with an arbitrary number of loops still contribute and loops correspond to orders in $G_N$.

\par
The classical limit of amplitudes is discussed in detail in Refs.~\cite{Cheung:2020gyp,Kosower:2018adc}.
In this limit, namely $S/\hbar \rightarrow \infty$, classical long-range contributions come from the part of the amplitude where graviton momenta are sent to zero.
This applies both to external momenta such as $q^\mu$ in Fig.~\ref{fig:amplitude1} and internal loop momenta.
On the other hand, scalar momenta are kept finite and the massive scalar is interpreted as a point particle.
\par
For the vertex function shown in Fig.~\ref{fig:amplitude1} we put the incoming scalar momentum on-shell so that ${k^2=m^2}$.
The amplitude is then multiplied together with a delta-funciton $\delta(kq)$ which, in the classical limit, puts the outgoing scalar on-shell as well (since we can neglect $q^2$ in $(k-q)^2$).
Note that, in the classical limit, the amplitude together with the delta-function is invariant if we translate $k^\mu$ with $q^\mu$ and we could also choose a symmetrical labelling if we let $k^\mu\rightarrow k^\mu + q^\mu/2$.
\par
The Lorentz covariance of the perturbative quantum field theoretic framework invites us to work in an arbitrary inertial frame.
It will be convenient then to introduce a notation which separates tensors into parallel and orthogonal parts with respect to the point particle momentum $k^\mu$.
We introduce the following projection operators
\begin{subequations}
  \label{eqn:lor1}
  \begin{eqnarray}
    &&\etat{\mn}=\frac{k_\mu k_\nu}{m^2}
    \ ,
    \\
    &&\etar{\mn}=\eta_\mn-\frac{k_\mu k_\nu}{m^2}
    \ ,
  \end{eqnarray}
\end{subequations}
and use similar symbols to signify projection with respect to these.
This notation is similar to that in \cite{Collado:2018isu}.
These operators are particularly simple in the inertial frame of $k^\mu$ where they are diagonal and represent the time- and space components of $\eta_\mn$ respectively.
Our definition of the Fourier transform between position and momentum space will be that of relativistic quantum field theory.
\par
In the classical limit we interpret the amplitude in Fig.~\ref{fig:amplitude1} as the source of the metric, $h_\mn$, generated from the point particle and the surrounding gravitational field.
The source of $h_\mn$, that is $\mathcal{M}_{\text{vertex}}^\mn$, is combined of an energy-momentum pseudo-tensor and a gauge fixing term.
From a comparison with the classical equations of motion we find that in the classical limit,
\begin{equation}
  2\pi\delta(kq)
  \ \mathcal{M}_{\text{vertex}}^\mn
  = - \kappa
  \ \tilde \tau^\mn
  - \frac{1}{\xi}
  \frac{4}{\kappa} \tilde H^\mn_{\text{non-linear}}
  \ ,
  \label{eqn:amp3}
\end{equation}
where $\tilde \tau^\mn$ and $\tilde H^\mn_{\text{non-linear}}$ are independent of $\xi$ but both depend on $q^\mu$ and $k^\mu$.
\par
The tensor $\tilde\tau^\mn$ is the total energy-momentum tensor of matter and gravitation in momentum space and is e.g. discussed in \cite{Weinberg:1972kfs,BjerrumBohr:2002ks}.
It is locally conserved and therefore obeys ${q_\mu \tilde \tau^\mn = 0}$.
To zeroth order, it is given by the point particle energy-momentum tensor of special relativity, and loop-corrections describe energy-momentum from the surrounding, self-interacting gravitational field.
The tensor $\tilde H^\mn_{\text{non-linear}}$ is the non-linear (in $h_\mn$) part of the gauge-breaking tensor, $H^\mn$, in momentum space.
\par
It is not clear that, in the classical limit, the dependence on $\xi$ of $\mathcal{M}_{\text{vertex}}^\mn$ can be reduced to that of the simple expression in Eq.~(\ref{eqn:amp3}), since each graviton self-interaction vertex includes a factor $\oov{\xi}$ and each graviton propagator a factor $\xi$.
However, due to several cancellations, all powers of $\xi$ different from $\oov{\xi}$ disappear in the classical limit.
\par
To get the metric, we solve the classical equation of motion ${\delta S=0}$ by contracting $\mathcal{M}^\mn_{\text{vertex}}$ with the graviton propagator Eq.~(\ref{eqn:pro1}).
Finally we can go to position space with a Fourier transform to get the \STM\ metric:
\begin{equation}
  g_\mn = \eta_\mn
  - \frac{\kappa}{2}
  \int \frac{d^Dq\ \delta(kq)\ e^{-iq x}}{(2\pi)^{D-1}}
  \frac{G_{\mn\ab}}{q^2}
  \mathcal{M}_{\text{vertex}}^\ab
  \ .
  \label{eqn:ext2}
\end{equation}
This exciting equation relates the metric from classical general relativity to the scattering amplitude, $\mathcal{M}_{\text{vertex}}^\ab$.
Although expected to hold to all orders in $G_N$ we only verify it to one-loop order in this article.
\par
While both the vertex function and the graviton propagator depend on $\xi$, the metric does not.
This is due to the Einstein field equations combined with the gauge condition $G_\sigma=0$.
If we separate the graviton propagator into two parts independent of $\xi$, $G_\ab^\mn = (G^{c}+\xi G^{gf})_\ab^\mn$, we find that the following combinations vanish, $G^{gf}\tilde \tau=0$ and $G^{c} \tilde H_{\text{non-linear}}=0$, where we have omitted indices.
These two equations correspond to the Einstein field equations and the gauge condition, respectively, and secure that $\xi$ disappears from the metric.
Using these relations, we get an expression for $h_\mn$ independent of $\xi$ in momentum space:
\begin{equation}
  \tilde h_\mn = \frac{\mathcal{P}^{-1}_{\mn\ab}}{q^2}
  \Big(
  \frac{\kappa^2}{2}
  \tilde \tau^\ab
  +2
  \tilde H_{\text{non-linear}}^\ab
  \Big)
  \ .
  \label{eqn:hmn1}
\end{equation}
In this equation and in Eq.~(\ref{eqn:ext2}) indices on the propagator were lowered with the flat space metric.
\par
Let us compare Eq.~(\ref{eqn:hmn1}) with the approach in~\cite{BjerrumBohr:2002ks}.
There, loop-corrections to $\tilde \tau^\mn$ was calculated in $D=4$ with the background field method and the metric was obtained in harmonic gauge by solving the classical Einstein field equations with the non-linear harmonic gauge condition ${\Gamma^\sigma_\mn g^\mn=0}$ which meant that a gauge-dependent term was added to the energy-momentum tensor.
In our approach the gauge-dependent term is already included in the amplitude in the form of $H_{\text{non-linear}}^\mn$ and this tensor exactly corresponds to their gauge-dependent correction to $\tilde\tau^\mn$.
\par
Eq.~\ref{eqn:hmn1} is particularly simple in \dDo\  gauge where ${\alpha=0}$.
In this gauge $\sqrt{-g}H^\mn$ is linear in $h_\mn$ which implies that $H_{\text{non-linear}}^\mn=0$ so that the second term on the right hand side disappears.
Thus in \dDo\  gauge, the graviton $h_\mn$ couples directly to the local energy-momentum tensor $\tau^\mn$.
In general the linear gauge of $\alpha=0$ is special since then, the $\xi$-dependence of the graviton self-interaction vertices disappears.
In this case ``Landau gauge'' $\xi\rightarrow0$ is possible.
\par
As an example we will first compute the tree-level contribution to $\mathcal{M}^\mn_{\text{vertex}}$ from which we derive the first order correction to the metric.
Afterwards we will focus on the one-loop contribution, where gravitational self-interactions first appear, which gives the $(G_N)^2$ metric contribution.
\subsection{Tree Level: Newton Potential in Arbitrary Dimensions}
\label{sec:newton}
As a simple example we compute the first order Newton correction to the \STM\ metric.
This comes from the tree diagram where a single graviton is connected to the scalar line.
We get, in the classical limit,
${i\mathcal{M}_{\text{tree}}^\mn = -i\kappa k^\mu k^\nu}$, 
where we have used the same labeling of momenta as in Fig.~\ref{fig:amplitude1} and the $h\phi^2$ vertex rule from Appendix~\ref{sec:frules} and neglected factors of $q^\mu$.
This amplitude is independent of the gauge parameters and for $\tilde\tau^\mn$ we find to zeroth order that
$\tilde \tau^\mn \approx 2\pi\delta(kq)k^\mu k^\nu$.
This is indeed conserved $q_\mu \tilde \tau^\mn=0$ and reproduces, in position space, the simple energy-momentum tensor of an inertial point particle.
Using Eq.~(\ref{eqn:ext2}) we propagate the tree-amplitude and go to position space to get the Newton potential in arbitrary dimensions:
\begin{equation}
  h^{(1)}_\mn 
  = -\frac{\mu}{\sqrt{-x_\bot^2}^{D-3}}
  \big(
  \etat{\mn} - \frac{1}{D-3} \etar{\mn}
  \big)
  \ .
  \label{eqn:res1}
\end{equation}
We use the Lorentz covariant notation of Eqs.~(\ref{eqn:lor1}).
The \STM\ parameter $\mu$ is
\begin{equation}
  \mu = \frac{16 \pi G_N m}{(D-2) \Omega_{D-2}}
  \ ,
  \label{eqn:mup1}
\end{equation}
where $\Omega_{d-1}$ is the surface area of a sphere in d-dimensional space and $\Omega_{d} = \frac{2\sqrt{\pi}^{d+1}}{\Gamma((d+1)/2)}$.
The first order metric in Eq.~\ref{eqn:res1} agrees with the results in \cite{Emparan:2008eg,Collado:2018isu}.
It is independent of both $\xi$, as expected, and $\alpha$ since $\alpha$ only enters in the self-interaction vertices.
Also, as expected, it satisfies the gauge condition $G_\sigma=0$ to first order in $G_N$.
%
%
%
%
\subsection{One-Loop Contribution to the Metric}
\label{sec:second}
\begin{figure}[h]
  \centering
  \includegraphics[width=5cm]{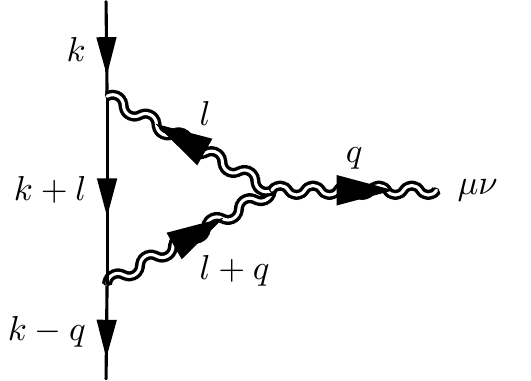}
  \caption{Feynman triangle diagram. The solid line is a massive scalar and wiggly lines are gravitons.}
  \label{fig:amplitude}
\end{figure}

The $(G_N)^2$ contribution to the metric comes from the triangle one-loop diagram in Fig.~\ref{fig:amplitude}.
Other one-loop diagrams do not contribute with non-analytic classical terms \cite{Bjerrum-Bohr:2018xdl}.
First, we will compute the one-loop Feynman diagram after which we can use Eq.~\eqreft{ext2} to derive the metric.
\par
Using the Feynman rules from Appendix~\ref{loopint} including the three-graviton vertex, $W_{h^3}^{\ab\ \gd\ \mn}(l,-l-q,q)$, and the graviton propagator presented in Eq.~\eqreft{pro1}, we get an expression for the amplitude:
\begin{eqnarray}
  &&
  2\pi\delta(kq) i \maM^\mn_\teve
  =
  2\pi\delta(kq)
  \frac{16m^2}{\kappa}
  \labelt{ext13}
  \\
  &&
  \qquad\qquad\quad
  \times
  \int
  \dDp{l}
  \frac{
    1
  }{l^2(l+q_\bot)^2\big((l+k)^2-m^2+i\epsilon\big)}
  \nonumber
  \\
  &&
  \qquad\qquad\quad
  \times
  \
  f_\ab
  \
  f_\gd
  \
  W_{h^3}^{\mn\ \ab\ \gd}(q,l,-l-q)
  \nonumber
  \ .
\end{eqnarray}
Here, we have introduced the tensor $f_\ab$ which describes the scalar-graviton vertices contracted with the graviton propagator:
\begin{equation}
  f_\ab
  =
  \frac{\kappa^2m}{2}
  \maP^{-1}_{\ab\rs}
  \frac{
    k^\rho
    k^\sigma
  }{
    m^2
  }
  \ .
\end{equation}
In this definition we have already made use of two simplifications due to the classical limit.
First, in the scalar-graviton vertices we can neglect graviton momenta in comparison to the scalar momentum so that both vertices are proportional to $k^\rho k^\sigma$.
Second, the momentum dependent part of the graviton propagator, $G_{\ab\rs}$, does not contribute and can be ignored so that we have instead, $\maP^{-1}_{\ab\rs}$.
This can be verified by analyzing the corresponding integrals but is understood intuitively since scalar-graviton vertices represent energy-momentum and are conserved, so that they disappear when they are contracted with the momentum from the propagator.
\par
Also, in Eq.~\eqreft{ext13} the $i\epsilon$-prescription should be included in the graviton propagators.
However, as it turns out it is only significant in the massive propagator.
\par
The triangle loop-integrals relevant for this amplitude have been treated in detail in \cite{Bjerrum-Bohr:2018xdl,Cristofoli:2020uzm} and are also discussed in Appendix~\ref{loopint}.
The three-graviton vertex introduces two graviton momenta in the numerator ($q_\mu$ or $l_\mu$) and we have then at most two loop momenta in the numerator.
\par
It is found in Appendix~\ref{loopint} that only the orthogonal part of the triangle diagrams contribute to the amplitude (orthogonal with respect to $k_\mu$).
This part can be reduced to a simple convolution integral which can easily be realized by rewriting the scalar propagator as follows:
\begin{subequations}
  \label{eqn:ext16}
  \begin{eqnarray}
    \frac{1}{(l+k)^2-m^2+i\epsilon}
    &=&
    \frac{1}{2kl+i\epsilon}
    \labelt{ext17}
    \\
    &=&
    \frac{1}{2kl}
    -
    i\frac{\pi}{2m}
    \delta(\frac{kl}{m})
    \labelt{ext17}
  \end{eqnarray}
\end{subequations}
In the first line we have neglected $l^2$ compared to $kl$ due to the classical limit and in the next line we separate the expression into real and imaginary parts.
It is then shown in Appendix~\ref{loopint} that only the imaginary part, i.e. the delta-function, contributes to the amplitude.
\par
In this way we get a simplified expression for the amplitude:
\begin{widetext}
  \begin{equation}
    2\pi\delta(kq)
    \maM^\mn_\teve
    =
    -
    \frac{4}{\kappa}
    \int \dDp{l}
    \frac{
      2\pi\delta(\frac{kl}{m})\  f_\ab
    }{
      l_\bot^2
    }
    \ 
    \frac{
      2\pi\delta(\frac{k(q-l)}{m})\  f_\gd
    }{
      (q_\bot-l_\bot)^2
    }
    \ 
    W_{h^3}^{\mn\ \ab\ \gd}(q_\bot,l_\bot,-l_\bot-q_\bot)
    \ .
    \labelt{ext3}
  \end{equation}
  Notice that the factors,
  \begin{equation}
    \frac{2\pi\delta(\frac{kl}{m}) f_\ab}{l_\bot^2}
    \ ,
    \frac{2\pi\delta(\frac{k(q-l)}{m}) f_\ab}{(q_\bot-l_\bot)^2}
    \ ,
  \end{equation}
  in the integrand correspond exactly to the first order metric, $h^{(1)}_\mn$, from Eq.~\eqreft{res1} in momentum space.
  The integral is then a convolution of two copies of $\tilde h^{(1)}_\mn$ together with the three-graviton vertex.
  Going to position space the convolution becomes simple multiplication and the two graviton momenta in the three-graviton vertex correspond to derivatives.
  In position space the integral in Eq.~\eqreft{ext3} then becomes a local, two derivatives, quadratic function of the first order metric (Eq.~\eqreft{res1}).
  This is no surprise and we know what this function is from Eq.~\eqreft{amp3}, i.e. the $h^2$ contribution to the local energy-momentum tensor, $\tau^\mn$, and $H^\mn$.
  \par
  We can now compute the amplitude.
  The triangle integrals have been reduced to bubble integrals and are given in Appendix~\ref{loopint} and the three-graviton vertex is given in Eqs.~\eqreft{ext4}.
  Using the definition of the amplitude in Eq.~\eqreft{amp3} we get:
  \begin{subequations}
    \label{eqn:amp4}
    \begin{eqnarray}
      \tilde \tau_{\text{1-loop}}^\mn
      =
      -&&2\pi\delta(\frac{kq}{m})
      \ \frac{\kappa^2 m^2 \Omega_{D-3}\sqrt{-q^2}^{D-3}}{64
        \cos(\frac{\pi}{2}D)
        (4\pi)^{D-3}}
      \bigg(
      \frac{D-7}{D-2} \eta^\mn_\parallel
      - \frac{(D-3)(3D-5)}{(D-2)^2}
      \Big(
      \eta_\bot^\mn - \frac{\ q^\mu q^\nu}{q^2}
      \Big)
      \bigg)
      \ ,
      \label{eqn:amp1}
      \\
      \tilde H_{\text{1-loop}}^\mn =
      \alpha\kappa^2
      \ &&2\pi\delta(\frac{kq}{m})
      \ \frac{\kappa^2 m^2 \Omega_{D-3}\sqrt{-q^2}^{D-3}}{64
        \cos(\frac{\pi}{2}D)
        (4\pi)^{D-3}}
      \frac{D-3}{D-2}
      \maP^{\mn}_\rs
      \frac{\ q^\rho q^\sigma}{q^2}
      \ .
      \label{eqn:amp2}
    \end{eqnarray}
  \end{subequations}
  These expressions are Lorentz covariant and valid in any dimension except for the factor $\cos(\frac{\pi}{2}D)$ in the denominator which makes them diverge in odd space-time dimensions.
  In principle we would then have to renormalize Eqs.~\eqreft{amp4} in odd $D$.
  However, when $D>5$ the divergent term is analytic in $q_\mu$ and can be ignored in the long-range classical limit since it describes local effects.
  In $D=5$ the divergent term has a non-analytic piece and a logarithm with an arbitrary scale appears in the metric in position space.
  Note, that this discussion concerns Eqs.~\eqreft{amp4} after dividing by $q^2$ to get the metric in momentum space.
  The $G^2_N$ contributions to $\tau^\mn$ and $H^\mn$ in Eqs.~\eqreft{amp4} have no divergent non-analytic pieces.
  We will discuss the divergences in detail in Sec.~\ref{sec:logarithm}.
  \par
  Clearly, $\tilde \tau_{\text{1-loop}}^\mn$ is locally conserved which implies that ${G_{gf}\tilde\tau_{\text{1-loop}}}$ vanishes as expected from the discussion above Eq.~(\ref{eqn:hmn1}).
  It is a straightforward check that ${G_{c}\tilde H_{\text{1-loop}}}$ disappears as well.
  This verifies that the metric is independent of $\xi$ to second order in $G_N$.
  At one-loop order $H^\mn$ is linear in $\alpha$ while $\tau^\mn$ is independent of $\alpha$.
  Going to higher orders in $G_N$ we would expect $\alpha$ to appear to any integer power in both $H^\mn$ and $\tau^\mn$.
  \par
  The metric in position space is computed with Eq.~\eqreft{ext2}.
  A useful formula for the relevant Fourier integrals is,
  \begin{equation}
    \int \frac{d^dq_\bot}{(2\pi)^d} \ e^{-ix_\bot q_\bot}
    (-q_\bot^2)^{\frac{n}{2}} =
    \frac{2^n}{\sqrt{\pi}^d}
    \frac{\Gamma(\frac{d+n}{2})}{\Gamma(-\frac{n}{2})}
    \frac{1}{(-x_\bot^2)^\frac{d+n}{2}}
    \ ,
    \label{eqn:int1}
  \end{equation}
  which can be found in e.g. Ref.~\cite{Collado:2018isu} and which was also used in Sec.~\ref{sec:newton}.
  Using this integral and Eqs.~\eqreft{amp4} we can go to position space for all $D\neq5$ and for the one-loop contribution to the metric we get:
  \begin{eqnarray}
    h^{(2)}_\mn =
    \frac{\mu^2}{r^{2(D-3)}}
    \Bigg(
    \oov{2} \etat{\mn}
    -
    \frac{
      (4\alpha-3)D
      -8\alpha + 5}
         {4(D-5)} \Xmnb{\mu}{\nu}
         -
         \frac{
           2(1-\alpha)D^2
           -(13-10\alpha)D
           +25-12\alpha
         }
              {4(D-3)^2(D-5)}
              \etar{\mn}
              \Bigg)
              \ .
              \label{eqn:hmn2}
  \end{eqnarray}
\end{widetext}
Here, $r^2=-x_\bot^2$.
The pole in $D=5$ makes it evident that, in this dimension, the amplitude was not regularized correctly in momentum space.
As expected, this metric satisfies $G_\sigma=0$ to second order in $G_N$ (see e.g. Eq.~(\ref{eqn:gau4}) where $G_\sigma$ is expanded to second order in $G_N$).
\par
In \dDo\  gauge where $\alpha=0$ we find agreement of Eq.~(\ref{eqn:hmn2}) with \cite{Collado:2018isu} in any dimension\ \footnote{Note, that there is a misprint in the fourth line of their Eq.~($5.34$) where $(D-p-3)^2$ should be replaced by $(D-p-3)$. We thank Paolo Di Vecchia for confirming this.}.
For harmonic gauge $\alpha=1$ we know only of any comparison in $D=4$ e.g. \cite{Weinberg:1972kfs}.
For general $\alpha$ we have compared Eq.~\eqreft{hmn2} with a derivation in Appendix~\ref{sec:GR} using only methods from classical general relativity and we find agreement.
\par
We can choose any value for $\alpha$ and we can e.g. use this freedom to remove the coefficient of $\etar{\mn}$.
The special choice of $\alpha=\frac{5}{6}$ removes the pole in $D=5$, which will be explained in the next section.
\section{Appearance of Logarithms in the Perturbative Expansion}
\label{sec:logarithm}
In this section we will focus on the divergences of Eqs.~(\ref{eqn:amp4}) and how this leads to a logarithmic term in the metric in $D=5$.
We will explain why this term appears and learn that besides $D=5$ logarithmic terms are only expected in $D=4$.
\par
The divergence in Eqs.~(\ref{eqn:amp4}) comes from the factor $\cos(\frac{\pi}{2}D)$ in the denominator in odd dimensions.
To analyze these divergences we use the dimensional regularization scheme and take the limit where the dimension goes near odd integer values, so that $D=5+2n+2\epsilon$ where $n$ is an integer and $\epsilon$ is infinitesimal.
After multiplying Eqs.~\eqreft{amp4} by the propagator to get the metric, the pole in $\epsilon$ depends on $q^\mu$ as $(-q^2)^n$ or:
\begin{equation}
  (-q^2)^{n-1}q_\sigma q_\rho
  \ .
  \labelt{ext5}
\end{equation}
These divergencies are analytic in all dimensions $D\neq5$ and here they describe local effects.
In $D=5$ (i.e. $n=0$), the term in Eq.~\eqreft{ext5} has a classical (non-analytic) piece and the renormalization process will introduce an arbitrary parameter in the metric.
\par
In all odd dimensions, after renormalization, the divergent term is removed and a finite classical dependence on $q^\mu$ remains.
This results in the following prescription in odd dimensions,
\begin{equation}
  \frac{\sqrt{-q^2}^{D-3}}{\cos(\frac{\pi}{2}D)}
  \rightarrow
  (-1)^\frac{D-3}{2}
  \frac{(-q^2)^\frac{D-3}{2}
    \ln(-r_0^2q^2)}{\pi}
  \ ,
  \label{eqn:rep1}
\end{equation}
where $r_0$ is an arbitrary scale which is introduced from the dimensional dependence of $G_N$.
With this replacement Eqs.~(\ref{eqn:amp4}) are finite in all dimensions.
\par
In principle it would be necessary to separate Eqs.~(\ref{eqn:amp4}) into two expressions for even/odd dimensions before going to position space.
However, all cases but $D=5$ can be treated simultaneously because the analytic functions can be neglected with dimensionally regularized integrals such as Eq.~(\ref{eqn:int1}) which gives the Fourier transform of $(-q_\bot^2)^{n/2}$.
In this equation it is seen that when $\frac{n}{2}$ is an integer so that we are transforming an analytic function, the result is zero (i.e. there is no long-range contribution although there would be a non-zero local part).
\par
In Eq.~(\ref{eqn:hmn2}) the pole in $D=5$ comes from the integral,
\begin{widetext}
  \begin{equation}
    \int \frac{d^{D}q \ \delta(\frac{kq}{m}) \ e^{-iq x}}{(2\pi)^{D-1}}
    \frac{\Omega_{D-3} \sqrt{-q^2}^{D-5}}{(4\pi)^{D-3}\cos(\frac{\pi}{2}D)}
    \frac{q_\mu q_\nu}{q^2}
    = 
    -\Big(
    \frac{2}{\Omega_{D-2} (D-3) \sqrt{-x_\bot^2}^{D-3}}
    \Big)^2
    \frac{1}{D-5}
    \Big(
    \etar{\mn} - 2(D-3)\frac{x^\bot_\mu x^\bot_\nu}{x_\bot^2}
    \Big)
    \ ,
    \label{eqn:int2}
  \end{equation}
  which was not regularized in momentum space to remove the divergence.
  We will now compute this integral in $D=5$ by using the replacement rule in Eq.~(\ref{eqn:rep1}).
  \par
  The logarithmic dependence on $q^2$ in Eq.~(\ref{eqn:rep1}) can be rewritten in terms of powers of $q^2$ with
  \begin{equation}
    \ln(-q^2)
    = \frac{1}{\epsilon}
    \Big(
    (-q^2)^{\epsilon}-1
    \Big)
    \ ,
    \label{eqn:ext1}
  \end{equation}
  where $\epsilon$ is infinitesimal.
  The Fourier integral Eq.~(\ref{eqn:int1}) can now be used.
  Using these tools, we get that in $D=5$ the integral corresponding to Eq.~(\ref{eqn:int2}) becomes:
  \begin{eqnarray}
    \int \frac{d^5q \ \delta(\frac{kq}{m}) \ e^{-iqx}}{(2\pi)^4}
    \ln(-r_0^2 q^2)
    \frac{q_\mu q_\nu}{q^2}
    =
    \frac{1}{2\pi^2\sqrt{-x_\bot^2}^{4}}
    \bigg(
    \etar{\mn}
    -6 \Xmnb{\mu}{\nu}
    -\Big(
    \etar{\mn} - 4\Xmnb{\mu}{\nu}
    \Big)
    \ln(-\frac{x_\bot^2 e^{2\gamma}}{4r_0^2})
    \bigg)
    \ .
    \label{eqn:int3}
  \end{eqnarray}
  Here $\gamma$ is the Euler-Mascheroni constant which can be removed by a redefinition of the scale, $r_0$.
  This integral is responsible for the appearance of a logarithmic dependence on the radial variable in $D=5$.
  \par
  Using Eq.~(\ref{eqn:int3}) we can compute the second order metric in $D=5$.
  After a redefinition of $r_0$ we get:
  \begin{eqnarray}
    h^{(2)}_\mn =
    \frac{\mu^2}{r^{4}}
    \Big(
    \oov{2} \etat{\mn}
    - \frac{2(6\alpha-5)\ln\frac{r}{r_0}-1}{16} \etar{\mn}
    + \frac{(6\alpha-5)(4\ln\frac{r}{r_0}-1)}{8} \Xmnb{\mu}{\nu}
    \Big)
    \ .
    \label{eqn:hmn3}
  \end{eqnarray}
\end{widetext}
Again, $r^2 = -x_\bot^2$.
We have not found this result in earlier literature, although a similar situation occurs in $D=4$ at third order in $G_N$ in \dDo\  gauge \cite{Goldberger:2004jt}.
In both cases a logarithmic dependence on the radial coordinate appears.
We will see that exactly in these two cases, this is expected, and that even to higher orders in $G_N$ we would not expect logarithms to appear in $D\geq6$.
The metric in Eq.~\eqreft{hmn3} is in agreement with the classical derivation in Appendix~\ref{sec:GR}.
\par
Note that we can make the logarithm disappear with the special choice $\alpha=\frac{5}{6}$.
The arbitrary scale, however, would in principle still be there.
In analogy, in $D=4$ we know that for $\alpha=1$ in harmonic gauge there is no logarithms.
\par
The arbitrary scale corresponds to a redundant gauge freedom.
It is well known from linearized gravity that even after choosing \dDo\  gauge, we can still translate the coordinates with $\epsilon^\sigma$ as long as it is a harmonic function, $\partial^2 \epsilon^\sigma=0$.
In our situation the relevant coordinate transformation is
\begin{equation}
  x^\mu
  \rightarrow
  x^\mu
  + \beta
  \frac{\mu^2}{r^4}
  x_\bot^\mu
  + \ ...
  \label{eqn:cor1}
\end{equation}
which does not change our gauge since ${(x_\bot^\sigma/r^4)}$ is a harmonic function (when $r\neq0$).
At higher orders in $G_N$ this coordinate transformation gets corrected which is indicated by the ellipsis.
Choosing $\beta$ in Eq.~(\ref{eqn:cor1}) appropriately makes the coordinate transformation equivalent to a scaling of the arbitrary parameter $r_0 \rightarrow \gamma r_0$ in Eq.~(\ref{eqn:hmn3}).
Thus, the arbitrary parameter in the metric in $D=5$ is unproblematic and is related to the coordinate system not being completely specified yet.
\par
In arbitrary space-time dimensions the equivalent transformation would be ${x_\bot^\sigma/r^{D-1}}$ which is a harmonic function (again, when $r\neq0$).
In any dimension we would be able to introduce an arbitrary parameter with such a transformation.
However, only in $D=4$ or $D=5$ does this lead to the appearance of logarithms in the metric.
This is due to the fact that only in these dimensions, such a transformation would be possible using the dimensions of $\mu$.
In $D=5$ it is accompanied by $\mu^2$ while in $D=4$ we get $\mu^3$ so that the logarithms appear respectively at second and third order.
In other dimensions the transformation would have to be scaled by fractional powers of $\mu$.
\section{Concluding Remarks}
We have analyzed the problem of deriving the \STM\ metric from scattering amplitudes in detail.
We calculated the metric to second order in $G_N$ and verified our general conclusions to this order.
These include that the metric is independent of the covariant parameter $\xi$, that it obeys the classical gauge condition $G_\sigma=0$, and that it simply is the Fourier transform of the three-point vertex of a scalar interacting with a graviton after propagation by the graviton propagator.
\par
In $D=5$ a logarithmic dependence appeared in position space analogous to the case in $D=4$ at third order in $G_N$ \cite{Goldberger:2004jt}.
We analyzed this curious phenomenon in terms of redundant gauge freedom and coordinate transformations.
This freedom makes it possible to introduce an arbitrary parameter in any dimension, though only in the two cases $D=4$ and $D=5$ does it lead to logarithmic terms in the metric.
\par
The full all-order expansion of the \STM\ metric from scattering amplitudes is still to be performed explicitly.
This requires an inductive relation between the loop amplitudes at different orders.
Already several exciting simplifications are known \cite{Bjerrum-Bohr:2018xdl}.
A logical continuation is to analyze the analogous problem for particles with spin and eventually look at quantum corrections \cite{Chung:2019yfs,BjerrumBohr:2002ks}.
Also, it would be interesting to continue investigations into solutions in classical general relativity in perturbative gauges such as the \dDo\ and harmonic gauges.
\begin{acknowledgments}
  I want to thank Poul Henrik Damgaard and Emil Bjerrum-Bohr for helpful discussions and comments.
  Also, I want to thank the anonymous referee for very helpful comments and for pointing out a weakness in an earlier argument regarding the classical equations of motion and their dependence on $\xi$.
\end{acknowledgments}

\appendix
\section{Feynman Rules}
\label{sec:frules}
The Feynman rules in covariant \dDt\ gauge are derived in similar fashion as other gauge fixing procedures such as the background field method in $D=4$ \cite{BjerrumBohr:2002kt,Donoghue:1995cz} and supergravity in \dDo\ gauge in arbitrary dimensions \cite{Collado:2018isu}. 
\par
We use the path integral method and expand the metric around flat space-time ${g_\mn = \eta_\mn + \kappa h_\mn}$.
We raise and lower all indices in this section with $\eta^\mn=\eta_\mn$ and use the tensors ${I^\mn_\ab}$ and ${\mathcal{P}^\mn_\ab}$ introduced under Eqs.~\eqreft{ext6}.
The terms in the action which are relevant for the one-loop computation are $h^2$, $\phi^2$, $h\phi^2$ and $h^3$.
\par
We expand the action from Eq.~\eqreft{act2} in powers of the fields:
\begin{equation}
  S = S_{h^2} + S_{\phi^2} + S_{h \phi^2} + S_{h^3} + ...
  \label{eqn:act1}
\end{equation}
The terms, $S_{h^2}$, $S_{\phi^2}$ and $S_{h \phi^2}$, are relatively simple and can be written conveniently in position space:
\begin{widetext}
\begin{subequations}
  \begin{eqnarray}
    S_{h^2}
    &=&
    \frac{1}{2}
    \int d^\ddi x
    \ h_{\mn}^{,\rho}
    \Big(
    \delta^\rho_\sigma \mathcal{P}^\mn_\ab
    -2(1-\oov{\xi}) \mathcal{P}^{\mn}_{\rho\kappa} \mathcal{P}_{\ab}^{\sigma\kappa}
    \Big)
    h^{\ab}_{,\sigma}
    \ ,
    \labelt{ext10}
    \\
    S_{\phi^2}
    &=&
    \frac{1}{2} \int d^\ddi x
    \Big(
    \phi^{,\nu} \phi_{,\nu}
    - m^2 \phi^2
    \Big)
    \ ,
    \\
    S_{h\phi^2}
    &=&
    -\frac{\kappa}{2}
    \int d^\ddi x
    \Big(
    h^\mn \phi_{,\mu} \phi_{,\nu}
    - \frac{1}{2} h^\mu_\mu (\phi^{,\nu} \phi_{,\nu} - m^2\phi^2)
    \Big)
    \ .
    \labelt{ext9}
  \end{eqnarray}
\end{subequations}
The three-graviton term is more complicated and we will write it in momentum space:
\begin{eqnarray}
  &&
  S_{h^3}
  =
  -\frac{2\kappa}{3}
  \int \dDp{q}
  \dDp{l_\teon}
  \dDp{l_\tetw}
  (2\pi)^D \delta^D(q+l_\teon+l_\tetw)
  W^{\mn\ \ab\ \gd}(q,l_\teon,l_\tetw)
  \tilde h_\mn^{(q)}
  \tilde h_\ab^{(1)}
  \tilde h_\gd^{(2)}
  \labelt{ext8}
\end{eqnarray}
Here, the superscripts on the gravitons indicate their dependence on momenta.
From the tensor, $W^{\mn\ \ab\ \gd}(q,l_\teon,l_\tetw)$, we can derive the three-graviton vertex rule.
We separate the tensor in two parts,
\begin{subequations}
  \labelt{ext4}
  \begin{equation}
    W^{\mn\ \ab\ \gd}
    =
    W_\tecl^{\mn\ \ab\ \gd}
    +
    \frac{1}{\xi}
    W_\tegf^{\mn\ \ab\ \gd}
    \ ,
  \end{equation}
  where, in this expression, we have hidden the momentum dependence of $W^{\mn\ \ab\ \gd}$.
  We find it convenient to write out the definition of $W^{\mn\ \ab\ \gd}$, when it is contracted with two gravitons as follows:
    \begin{eqnarray}
      &&
      W_\tecl^{\mn\ \ab\ \gd}(q,l_\teon,l_\tetw)
      h^\teon_\ab
      h^\tetw_\gd
      =
      -\frac{1}{2}
      Q^{\mn\ \ez\ \et}
      \eta^\rs
      \Gamma^{\kappa\ab}_{\rho\ez}
      \Gamma^{\lambda\gd}_{\sigma\et}
      l^\teon_\kappa l^\tetw_\lambda
      h^\teon_\ab
      h^\tetw_\gd
      \labelt{ext15}
      \\
      &&
      \hspace{3cm}
      +
      \frac{1}{4}
      \Big(
      I^\mn_\ab I^\gd_\rs
      +
      I^\mn_\rs I^\gd_\ab
      -
      2
      I^\mn_{\kappa\zeta}
      I^{\zeta\eta}_\ab
      I^\gd_{\eta\theta}
      I^{\theta\kappa}_\rs
      \Big)
      l_\teon^\rho
      l_\teon^\sigma
      h_\teon^\ab
      h^\tetw_\gd
      \ \ + \ \ 
      \langle 1\leftrightarrow2 \rangle
      \nonumber{}
      \\
      &&
      \hspace{3cm}
      +
      \frac{1}{2}
      \Big(
      \maP^\mn_{\gamma\kappa}
      \eta^{\kappa\theta}
      \maPi^\rs_{\delta\theta}
      +
      \frac{1}{2(D-2)}
      \eta^\rs
      I^\mn_\gd
      -
      \frac{1}{4}
      \eta_\gd
      I^{\mn\rs}
      \Big)
      Q_{\rs\ \ab\ \pe}
      l_\teon^\phi
      l_\teon^\epsilon
      h_\teon^\ab
      h_\tetw^\gd
      \ \ + \ \ 
      \langle 1\leftrightarrow2 \rangle
      \nonumber{}
    \end{eqnarray}
    \begin{eqnarray}
      W_\tegf^{\mn\ \ab\ \gd}(q,l_\teon,l_\tetw)
      h^\teon_\ab
      h^\tetw_\gd
      &=&
      -\frac{1}{2}\alpha
      \maP^{\mn\rs}
      \Gamma_{\rho\gd}^{\kappa\ab}
      q_\sigma
      l_\kappa^\teon
      h_\ab^\teon
      h^\gd_\tetw
      \ \ + \ \ 
      \langle 1\leftrightarrow2 \rangle
      \labelt{ext14}
      \\
      &&
      -\frac{1}{2}
      \alpha
      \Big(
      \Gamma^{\rho\mn}_{\kappa\ab}
      l_\tetw^\kappa
      h^\gd_\tetw
      +
      \eta^{\lambda\rho}
      \Gamma^{\kappa\mn}_{\lambda\gd}
      q_\kappa
      h_\tetw^\gd
      \Big)
      \maP^\ab_\rs
      h^\teon_\ab
      l^\sigma_\teon
      \ \ + \ \ 
      \langle 1\leftrightarrow2 \rangle
      \nonumber{}
    \end{eqnarray}
\end{subequations}
\end{widetext}
The notation, $\langle 1\leftrightarrow2 \rangle$, means that a term should be added where the subscripts (or superscripts) $1$ and $2$ are interchanged (i.e. on the gravitons and momenta).
After these terms are added the tensors $W_\tecl^{\mn\ \ab\ \gd}$ and $W_\tegf^{\mn\ \ab\ \gd}$ are simply read off by removing the graviton fields.
When conservation of momentum is enforced, i.e. $q=-l_\teon-l_\tetw$, the tensors are symmetric in the three graviton indices and their momenta.
In Eqs.~\eqreft{ext4} we have introduced two tensors:
\begin{subequations}
  \begin{eqnarray}
    &&
    Q^{\mn\ \ab\ \gd}
    =
    \eta^\mn \maP^{\ab\gd}
    -2
    I^{\mn\rs}
    \maP^{\ab}_{\rho\phi}
    \eta^\pe
    \maP_{\epsilon\sigma}^{\gd}
    \labelt{ext7}
    \\
    &&
    \Gamma^{\rho\ab}_{\sigma\gd}
    =
    I^\ab_{\sigma\kappa}I^{\kappa\rho}_\gd
    -
    \frac{1}{2}
    I^\ab_\gd \delta^\rho_\sigma
  \end{eqnarray}
\end{subequations}
Note that $Q^{\mn\ \ab\ \gd}$ is symmetric when any pair of indices is exchanged with any other i.e. $\mn \leftrightarrow \ab$ although this is not evident from its definition in Eq.~\eqreft{ext7}.
\par
When Eqs.~\eqreft{ext4} are used for the amplitude computation in Sec.~\ref{sec:second} the third line of Eq.~\eqreft{ext15} and the second line of Eq.~\eqreft{ext14} do not contribute.
This is related to the Einstein field equations and the gauge condition $G_\sigma=0$ respectively.
\par
The three-graviton vertex rule is derived from Eq.~\eqreft{ext8} and the $\phi^2 h$ vertex from Eq.~\eqreft{ext9}:
\begin{figure}[H]
  \raggedright
  \begin{tikzpicture}
    \node
    (gravQubic)
        {
          \includegraphics{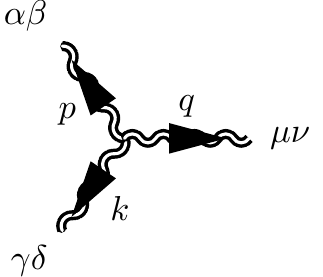}
        };
        \node
            [text width=5cm,right=-.1cm of gravQubic]
            {
              $=-i4\kappa W^{\mn\ \ab\ \gd}(q,p,k)$
            };
  \end{tikzpicture}
\end{figure}
\begin{figure}[H]
  \raggedright
  \begin{tikzpicture}
    \node
    (scalar-grav)
        {
          \includegraphics{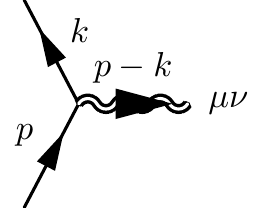}
        };
        \node
            [right=0cm of scalar-grav]
            {
              $= \ -i\frac{\kappa}{2} \big(p^\mu k^\nu + k^\mu p^\nu - \eta^\mn(pk-m^2)\big)$
            };
  \end{tikzpicture}
\end{figure}
We get the graviton propagator by inverting the quadratic operator in the $h^2$ term of the action in Eq.~\eqreft{ext10}.
In momentum space, this term reads
\begin{equation}
  S_{h^2} =
  \frac{1}{2}
  \int \dba{l}{D}
  \tilde h_\mn ^{\ \dagger}
  \ l^2\ \Delta^\mn_\ab
  \ \tilde h^\ab
  \ ,
\end{equation}
where the dagger signifies complex conjugation and
\begin{equation}
  \Delta^\mn_\ab =
  \mathcal{P}^\mn_\ab - 2(1-\frac{1}{\xi})\mathcal{P}^{\mn}_{\rho\kappa} \frac{\ l^\rho l_\sigma}{l^2} \mathcal{P}^{\kappa \sigma}_{\ab}
  \ ,
  \labelt{ext11}
\end{equation}
is a tensor depending on both the momentum $l^\mu$ and the covariant gauge parameter $\xi$.
We invert the tensor and for $(\Delta^\mn_\ab)^{-1}$, we find (as in Eq.~\eqreft{pro1})
\begin{equation}
  G^\mn_\ab
  = {\mathcal{P}^{-1}}^\mn_\ab
  - 2(1-\xi)
  I^{\mn}_{\rho\kappa} \frac{\ l^\rho  l_\sigma}{l^2} I^{\kappa \sigma}_{\ab}
  \ ,
  \labelt{ext12}
\end{equation}
so that $G^\mn_\ab \Delta^\ab_\gd = I^\mn_\gd$.
Here, $\mathcal{P}^{-1}$ is the inverse operator to $\mathcal{P}$ defined in Eq~(\ref{eqn:ddo1}).
\par
The structure of these operators can be analyzed by separating them into parts that are independent of $\xi$, so that ${G = G_c + \xi G_{gf}}$ and ${\Delta = \Delta_c + \oov{\xi} \Delta_{gf}}$, where we have omitted indices.
These parts can be read off from Eqs.~\eqreft{ext11} and~\eqreft{ext12}.
They obey simple idendities,
\begin{subequations}
  \begin{eqnarray}
    &&\Delta_c \  G_{gf} = \Delta_{gf} \  G_c = 0
    \ ,
    \\
    &&\Delta \  G = \Delta_c \  G_c + \Delta_{gf} \  G_{gf} = I
    \ ,
  \end{eqnarray}
\end{subequations}
in which we have left out indices, but matrix multiplication is understood.
\par
The graviton propagator is then
\begin{equation}
  \frac{i}{l^2+i\epsilon} G^\mn_\ab
  \ ,
\end{equation}
and the scalar propagator $\frac{i}{l^2-m^2+i\epsilon}$.
\section{Triangle Loop Integrals}
\label{loopint}
The integral for the one-loop amplitude in Sec.~\ref{sec:second} is:
\begin{equation}
  \int
  \dDp{l}
  \frac{
    f_\ab
    f_\gd
    W_{h^3}^{\mn\ \ab\ \gd}(q,l,-l-q)
  }{l^2(l+q_\bot)^2\big((l+k)^2-m^2+i\epsilon\big)}
  \ .
  \labelt{ext19}
\end{equation}
As discussed there, in the classical limit, the massive propagator can be written as:
\begin{equation}
  \frac{1}{(l+k)^2-m^2+i\epsilon}
  =
  \frac{1}{2k l_\pll}
  -
  i\frac{\pi}{2m}
  \delta(\frac{k l_\pll}{m})
  \labelt{ext18}
\end{equation}
We want to show that only the imaginary part of Eq.~\eqreft{ext18} contributes to the amplitude.
In this equation, the imaginary part is an even function of $l_\pll^\mu$ while the real part is an odd function.
We will then show that, ignoring the massive propagator, the remaining integrand in Eq.~\eqreft{ext19} is even in $l_\pll^\mu$.
\par
We use Eqs.~\eqreft{ext4} for the definition of $W^{\mn\ \ab\ \gd}$ where $h^\teon_\ab$ and $h^\tetw_\gd$ both correspond to $f_\rs$ and $l^\teon$ corresponds to $l$ and $l^\tetw$ to $-l-q_\bot$.
There are three combinations of momenta that occur in the numerator of Eq.~\eqreft{ext19}, namely:
\begin{equation}
  (l_\teon^\mu+l_\tetw^\mu)q^\nu_\bot
  \ ,\ 
  l_\teon^\mu l_\tetw^\nu + l_\tetw^\mu l_\teon^\nu
  \ ,\
  l_\teon^\mu l_\teon^\nu + l_\tetw^\mu l_\tetw^\nu
  \labelt{ext20}
\end{equation}
First, we see that the graviton propagators are even in $l_\pll$.
This is certainly so for $l^2=l_\bot^2+l_\pll^2$ and also in the case:
\begin{equation}
  (l+q_\bot)^2 = (l_\bot+q_\bot)^2 + l_\pll^2
  \ .
\end{equation}
Here, it is important that the parallel part of $q$ is zero.
\par
For the first case of Eq.~\eqreft{ext20} the numerator is independent of $l_\mu$ since,
\begin{equation}
  (l_\teon^\mu+l_\tetw^\mu)q^\nu_\bot = -q^\mu_\bot q^\nu_\bot
  \ ,
\end{equation}
so that, indeed, the remaining integrand is even in $l_\pll$.
\par
The two other cases of Eq.~\eqreft{ext20} are similar:
\begin{eqnarray}
  &&
  \frac{
    l^\mu_\teon l^\nu_\tetw
    +
    l^\mu_\tetw l^\nu_\teon
  }{
    2
  }
  =
  -
  I^\mn_\rs
  l_\pll^\rho
  (q_\bot^\sigma+ 2 l_\bot^\sigma)
  +
  \big(
  \text{even in $l_\pll$}
  \big)
  \nonumber
  \\
  &&
  \frac{
    l^\mu_\teon l^\nu_\teon
    +
    l^\mu_\tetw l^\nu_\tetw
  }{
    2
  }
  =
  I^\mn_\rs
  l_\pll^\rho
  (q_\bot^\sigma+ 2 l_\bot^\sigma)
  +
  \big(
  \text{even in $l_\pll$}
  \big)
  \nonumber
\end{eqnarray}
It will now be shown that the odd part in $l_\pll$ of these expressions vanishes.
This corresponds to the integral of $q_\bot+2l_\bot$ which must be proportional to $q_\bot$.
We contract the integral with $q_\bot$:
\begin{subequations}
  \begin{eqnarray}
    q^\bot_\mu
    \int
    \dDp{l}
    \frac{
      (q_\bot^\mu + 2l_\bot^\mu) l^\nu_\pll
    }{
      l^2
      (l+q_\bot)^2
      \big((l+k)^2-m^2+i\epsilon\big)
    }
    \\
    =
    \int
    \dDp{l}
    \frac{
      \Big(
      (q_\bot+l)^2-l^2
      \Big)
      l^\nu_\pll
    }{
      l^2
      (l+q_\bot)^2
      \big((l+k)^2-m^2+i\epsilon\big)
    }
    \ .
  \end{eqnarray}
\end{subequations}
The integral in the second line does not have any classical piece and hence the integral vanishes in the classical limit.
\par
In this way we conclude that only the imaginary part of the massive propagator contributes to the amplitude.
For reference, we give the full expressions for the triangle integrals below although, as shown, only the orthogonal part contributes to the amplitude.

We define the triangle integrals as,
\begin{subequations}
  \begin{eqnarray}
    I
    &=&
    \int \dDp{l}
    \frac{1}{l^2
      (l+q_\bot)^2
      \big((l+k)^2-m^2+i\epsilon\big)}
    \\
    I^\mu
    &=&
    \int \dDp{l}
    \frac{l^\mu}{l^2
      (l+q_\bot)^2
      \big((l+k)^2-m^2+i\epsilon\big)}
    \\
    I^\mn
    &=&
    \int \dDp{l}
    \frac{l^\mu l^\nu}{
      l^2
      (l+q_\bot)^2
      \big((l+k)^2-m^2+i\epsilon\big)}
    \ \ \ \ 
  \end{eqnarray}
\end{subequations}
where in each expression the $i\epsilon$-prescription should also be included in the graviton propagators.
In the classical limit, they are given by:
\begin{subequations}
  \begin{eqnarray}
    I
    &=&
    -\frac{i}{4m} N_{D-1}
    \\
    I^\mu
    &=&
    -\frac{i}{4m} N^\mu_{D-1}
    +
    \frac{i}{2m^2} N_D \ k^\mu
    \\
    I^\mn
    &=&
    -\frac{i}{4m} N^\mn_{D-1}
    -
    \frac{i}{4m^2}
    N_D
    \Big(
    q^\mu_\bot k^\nu
    +
    k^\mu q^\nu_\bot
    \Big)
  \end{eqnarray}
\end{subequations}
The integral $N_{D-1}$ is given by,
\begin{equation}
  N_{D-1}
  =
  \int
  \dDp{l}
  2\pi \delta
  \Big(
  \frac{kl}{m}
  \Big)
  \frac{1}{l_\bot^2(l_\bot-q_\bot)^2}
  \ ,
\end{equation}
and the integrals $N^\mu_{D-1}$ and $N^\mn_{D-1}$ are given by the same expression with $l_\bot^\mu$ and $l_\bot^\mu l_\bot^\nu$ in the numerator respectively.
They are evaluated to:
\begin{subequations}
  \begin{eqnarray}
    &&
    N_{D-1}
    =
    \frac{
      \Omega_{D-3} \sqrt{-q_\bot^2}^{D-5}
    }{
      4 (4\pi)^{D-3}
      \cos(\frac{\pi}{2}D)
    }
    \labelt{eqn1}
    \\
    &&
    N^\mu_{D-1}
    =
    -\frac{N_{D-1}}{2}
    q_\bot^\mu
    \\
    &&
    N^\mn_{D-1}
    =
    \frac{q_\bot^2 \ N_{D-1}}{4(D-2)} 
    \Big(
    (D-1)
    \frac{
      q^\mu_\bot q^\nu_\bot
    }{
      q_\bot^2
    }
    -
    \eta^\mn_\bot
    \Big)
  \end{eqnarray}
\end{subequations}
The function $N_D$ is then defined by Eq.~\eqreft{eqn1} and is:
\begin{eqnarray}
  N_D
  =
  -
  \frac{
    \Omega_{D-2}
    \sqrt{-q_\bot^2}^{D-4}
  }{
    4(4\pi)^{D-2}
    \sin(\frac{\pi}{2} D)
  }
\end{eqnarray}
\section{Classical derivation of the \STM\ metric in \dDt\  coordinates}
\label{sec:GR}
We derive the \STM\ metric to second order in $G_N$ in \dDt\ coordinates that satisfy $G_\sigma=0$ with methods from classical general relativity.
We change coordinates from the standard \STM\ metric in spherical coordinates to new cartesian-like coordinates which we determine perturbatively to obey the \dDt\  gauge condition.
The method is analogous to that in Weinberg \cite{Weinberg:1972kfs} for harmonic gauge in $D=4$.
\par
In standard coordinates, the \STM\ metric is given by (see e.g. \cite{Emparan:2008eg}):
\begin{equation}
  d\tau^2 =
  (1-\frac{\mu}{R^\dmt}) dt^2
  -\frac{1}{1-\frac{\mu}{R^\dmt}} dR^2
  - R^2 d\Omega^2_{D-2}
  \label{eqn:met1}
  \ .
\end{equation}
Here $n=D-3$ and $\mu$ is the \STM\ parameter from Eq.~(\ref{eqn:mup1}).
The \STM\ metric solves the Einstein field equations in arbitrary dimensions $D$.
\par
We change the radial coordinate $R$ into a new one $r$ and then go to cartesian coordinates with respect to $r$.
We determine the relationship between $r$ and $R(r)$ perturbatively so that the cartesian-like coordinates obey the gauge condition $G_\sigma=0$.
The metric in terms of the new coordinates in the inertial frame of the point particle is
\begin{equation}
  d\tau^2 =
  B dt^2
  -\frac{1}{B} \frac{dR^2}{dr^2} (\frac{\vec{x}d\vec{x}}{r})^2
  - \frac{R^2}{r^2} \Big(  d\vec{x}^2 - (\frac{\vec{x}d\vec{x}}{r})^2  \Big)
  \ ,
\end{equation}
where $B=1-\frac{\mu}{R^n}$ and $r^2 = \absvec{x}^2$.
We generalize to the covariant notation of Eqs.~(\ref{eqn:lor1}) and get
\begin{equation}
  g_{\mn} =
  B \etat{\mn}
  +\frac{R^2}{r^2}\etar{\mn}
  +(\frac{1}{B}\frac{dR^2}{dr^2}-\frac{R^2}{r^2}) \Xmnb{\mu}{\nu}
  \ ,
  \label{eqn:met2}
\end{equation}
where now $r^2 = -x_\bot^2$.
We expand the \STM\ radial coordinate $R$ in terms of the new coordinate $r$ in powers of $\mu$:
\begin{equation}
  R = r
  \big(
  1 + a\fcr + b \ \Big(a\frac{\mu}{r^\dmt}\Big)^2 + ...
  \big)
  \ .
  \label{eqn:exp1}
\end{equation}
Where $a$ and $b$ are to be determined by the condition $G_\sigma=0$.
As we will see, this expansion is not sufficient in $D=5$ and the coefficient $b$ has to changed $b \rightarrow b_0+b_1 \ln\frac{r}{r_0}$.
For now we will ignore $D=5$ and continue.
Inserting our expansion of $R$ into the metric Eq.~(\ref{eqn:met2}) we get an expansion of the metric depending on the coefficients $a$ and $b$:
\begin{equation}
  g_\mn = \eta_\mn + \hnnm{1}{\mn} + \hnnm{2}{\mn} + ...
\end{equation}
The gauge fixing function is expanded similarly
\begin{eqnarray}
  &&G_\sigma \approx
  \hnn{1}^\mu_{\sigma,\mu} - \frac{1}{2} \hnn{1}_{\mu,\sigma}^\mu
  \label{eqn:gau4}
  \\
  &&+\hnn{2}^\mu_{\sigma,\mu} - \frac{1}{2} \hnn{2}_{\mu,\sigma}^\mu
  - \alpha\
  \Big(
  h_{(1)}^{\mn} h^{(1)}_{\sigma\mu,\nu}
  -\frac{1}{2} h_{(1)}^\mn h^{(1)}_{\mn,\sigma}
  \Big)
  \ ,
  \nonumber{}
\end{eqnarray}
where the first line is the first order term and the second line the second order term.
Perturbatively $G_\sigma=0$ means that each line vanishes by itself.
The first order term of $G_\sigma$ determines the coefficient $a$.
We compute $h_\mn^{(1)}$ in terms of $a$
\begin{equation}
  h^{(1)}_\mn = \frac{\mu}{r^\dmt} \Big(-\etat{\mn}
  +2a\etar{\mn} - \big(2na-1\big) \Xmnb{\mu}{\nu} \Big)
  \ ,
  \label{eqn:met3}
\end{equation}
from which we find the first order gauge condition
\begin{equation}
  \hnn{1}^\mu_{\sigma,\mu} - \frac{1}{2} \hnn{1}_{\mu,\sigma}^\mu = \frac{\mu}{r^{n+1}}(2na-1)
  \frac{x^\bot_\sigma}{r}
  \ ,
\end{equation}
so that $G_\sigma=0$ means $a=\oov{2n}$.
Eq.~(\ref{eqn:met3}) then agrees with our tree-level result.
\par
Going to second order we find an expression for $h^{(2)}_\mn$ in terms of $b$:
\begin{equation}
  h^{(2)}_\mn =
  \frac{\mu^2}{r^{2\dmt}}
  \big(\oov{2} \etat{\mn}
  + \frac{2b+1}{4\dmt^2} \etar{\mn}
  - \frac{4b+\dmt-2}{4\dmt} \Xmnb{\mu}{\nu}  \big)
  \ .
  \label{eqn:met4}
\end{equation}
The second order gauge condition reads:
\begin{subequations}
  \label{eqn:gau3}
  \begin{equation}
    \hnn{2}^\mu_{\sigma,\mu} - \frac{1}{2} \hnn{2}_{\mu,\sigma}^\mu
    = \alpha\
    \Big(
    h_{(1)}^{\mn} h^{(1)}_{\sigma\mu,\nu}
    -\frac{1}{2} h_{(1)}^\mn h^{(1)}_{\mn,\sigma}
    \Big)
    \ .
    \label{eqn:gau2}
  \end{equation}
  For the right hand side we find
  \begin{equation}
    h_{(1)}^{\mn} h^{(1)}_{\sigma\mu,\nu}
    -\frac{1}{2} h_{(1)}^\mn h^{(1)}_{\mn,\sigma}
    = -\frac{\mu^2}{r^{2\dmt+1}} \frac{\dmt+1}{2}
    \frac{x^\bot_\sigma}{r}
    \ ,
  \end{equation}
  and the left hand side:
  \begin{equation}
    \hnn{2}^\mu_{\sigma,\mu} - \frac{1}{2} \hnn{2}_{\mu,\sigma}^\mu
    =-\frac{\mu^2}{r^{2\dmt+1}} \frac{\dmt^2+1+(\dmt-2)b}{2\dmt}
    \frac{x^\bot_\sigma}{r}
    \ .
    \label{eqn:gau1}
  \end{equation}
\end{subequations}
Combining Eqs.~(\ref{eqn:gau3}) we determine $b$ to be:
\begin{equation}
  b = \frac{-(1-\alpha)\dmt^2+\alpha\dmt -1}{\dmt-2}
  \ .
\end{equation}
We see that $b$ diverges in $D=5$ which means our choice of expansion of $R(r)$ must be changed in $D=5$.
Inserting $b$ into $h^{(2)}_\mn$ in Eq.~(\ref{eqn:met4}) produces the same result as our one-loop computation for $D\neq5$.

\subsection{Appearance of a Logarithm in $D=5$}
In $D=5$ it is necessary to generalize the expansion of $R$ in terms of $r$.
We change Eq.~(\ref{eqn:exp1}) into
\begin{equation}
  R = r
  \big(
  1 + a\fcr + (b_0 + b_1 \ln\frac{r}{r_0}) \ \Big(a\frac{\mu}{r^\dmt}\Big)^2 + ...
  \big)
  \ ,
\end{equation}
where we have let $b\rightarrow b_0+b_1\ln\frac{r}{r_0}$.
We repeat the analogous steps as above with the new expansion of $R$.
For example Eq.~(\ref{eqn:gau1}) changes into
\begin{eqnarray}
  &&\hnn{2}^\mu_{\sigma,\mu} - \frac{1}{2} \hnn{2}_{\nu,\sigma}^\nu
  =
  -\frac{\mu^2}{r^{2\dmt+1}}
  \frac{x^\bot_\sigma}{r}
  \Big(
  \frac{\dmt^2+1+(\dmt-2)b}{2\dmt}
  \nonumber{}
  \\
  &&\qquad \qquad \qquad \qquad \qquad \qquad \qquad
  -\frac{3\dmt-2}{4\dmt^2}b_1
  \Big)
  \ ,
\end{eqnarray}
where $b=b_0+b_1 \ln\frac{r}{r_0}$.
The second order gauge condition Eq.~(\ref{eqn:gau2}) becomes:
\begin{equation}
  (\dmt-2)b-\frac{3\dmt-2}{2\dmt} b_1 = \alpha \dmt (\dmt+1) - \dmt^2 -1
\end{equation}
This equation is identical to the one that determined $b$ above in Eqs.~(\ref{eqn:gau3}) only that the term with $b_1$ is new and that $b$ now includes a logarithmic term.
For $D\neq5$ we are forced to remove the logarithmic dependence in $b$ so that $b_1=0$.
However, in $D=5$ we find that $b_1=5-6\alpha\neq0$ while both $b_0$ and $r_0$ are arbitrary.
We compute $h^{(2)}_\mn$ in terms of $b_0$ and $b_1$ for $D=5$
\begin{equation}
  h^{(2)}_\mn =
  \frac{\mu^2}{r^{4}}
  \big(\oov{2} \etat{\mn}
  + \frac{2b+1}{16} \etar{\mn}
  - \frac{4b-b_1}{8} \Xmnb{\mu}{\nu}  \big)
  \ ,
\end{equation}
where again $b=b_0+b_1 \ln\frac{r}{r_0}$.
Inserting $b_1=5-6\alpha$ and $b_0$, $r_0$ arbitrary produces the same result as our one-loop calculation for $D=5$.
\bibliography{refs}
\end{document}